\documentclass[runningheads]{llncs}

\usepackage{graphicx}
\usepackage{cmap}                                 
\usepackage{tabularx}                             
\usepackage{booktabs}                             

\usepackage{amssymb,amsmath}                      
\usepackage{hyperref}                             
\usepackage{flafter}                              
\usepackage[caption=false]{subfig}
\usepackage{url}                    
\usepackage{algorithmic}
\usepackage{algorithm}
\usepackage{todonotes}
\usepackage{marvosym}

\begin{document}
%
\title{Optimization of QKD Networks with Classical and Quantum Annealing}
\titlerunning{Optimization of QKD Networks with Classical and Quantum Annealing}
%
\author{Bob Godar\inst{1},
Christoph Roch\inst{1},
Jonas Stein\inst{1},
Marc Geitz \inst{2},
Bettina Lehmann \inst{2},
Matthias Gunkel \inst{3},
Volker Fürst \inst{3}\and
Fred Hofmann \inst{3}}
\authorrunning{B. Godar et al.}
%
\institute{Ludwig-Maximilians-Universität München, Geschwister-Scholl-Platz 1, 80539 München, Germany \\ \email{Christoph.Roch@ifi.lmu.de}\\ 
\and
T-Labs, Winterfeldtstraße 21, 10781 Berlin, Germany \\ \and
Deutsche Telekom Technik GmbH, Eschollbrücker Str. 12, 64283 Darmstadt, Germany\\
}
\maketitle              
\begin{abstract}
This paper analyses a classical and a quantum annealing approach to compute the minimum deployment of Quantum Key Distribution (QKD) hardware in a tier 1 provider network. The ensemble of QKD systems needs to be able to exchange as many encryption keys between all network nodes in order to encrypt the data payload, which is defined by traffic demand matrices. Redundancy and latency requirements add additional boundary conditions. The result of the optimization problem yields a classical heuristic network planners may utilize for planning future QKD quantum networks.

\end{abstract}

\section{Introduction}

Quantum Computers being able to execute Shor's algorithm \cite{shor1999polynomial} will be able to attack nowadays asymmetric cryptographic systems, like Diffie-Helman, Rivest Shamir Adleman (RSA) or Elliptic Curve Cryptography (ECC). As of today, a typical attack scenario is the harvesting of confidential network data traffic potentially followed by later decryption. Quantum computers are not yet ready to perform complex operations like factoring RSA 2048 bit keys due to the lacking number of qubits, qubit connectivity, gate fidelity and coherence time, but as these devices grow in capacity \cite{gambetta_2020}, the threat is increasing.

Since the business model of telecommunication providers depends strongly on the confidentiality of their services, there is a strong interest to investigate technology enabling quantum secure communication services. This paper focuses on Quantum Key Distribution (QKD) as technology option. QKD enables the secure distribution of encryption keys between two distant locations where the laws of quantum physics reveal possible eavesdroppers. The distributed encryption keys can then be used to encrypt the network payload using a quantum secure symmetric encryption protocol like Advanced Encryption Standard (AES). One AES encryption key is good for encrypting about 1 TBit of network data before the encryption key needs to be replaced \cite{dworkin2007sp}, requiring a constant exchange of encryption keys between the QKD systems.

The integration of QKD systems to provider networks requires key management, key storage and encryptor systems. Integration prototypes are being built in the course of the OpenQKD project, specifically in the Madrid or Berlin testbeds by Telefonica and Deutsche Telekom respectively \cite{Braun:21,Lopez:21}. These integration prototypes do not only go beyond single point to point QKD links, but try to deploy the quantum key exchange in connected networks. Due to the distance limitation of quantum optical links, which is about 100km due to fiber attenuation, these networks require so-called trusted intermediate network nodes (TN) to span larger distances \cite{xu2009field}. Due to the fact that quantum information needs to become classical within a TN, these are seen as the biggest security weakness of QKD networks. In the future, however, quantum repeater networks will solve this issue.

This paper focuses on the optimization of a QKD network. The work is based on anonymized network data of a Deutsche Telekom (DT) fiber optical network deployed in the field. It consists of 29 network nodes and 48 network connections. The quantum key distribution rate depends on the attenuation of the network links. Every network node is a source as well as a sink of network traffic. The encryption key distribution rate between all network nodes needs to be high enough to encrypt the network traffic to be transmitted, assuming AES encryption with 256 bit keys. In this paper, we show a method to derive the most effective deployment of QKD systems in provider networks and apply the method to the network of Deutsche Telekom.

We approach the problem of QKD network optimization by creation of a Quadratic Unconstrained Binary Optimization (QUBO) cost function to decide which network node is to be equipped with QKD servers. The QUBO can be minimized using a Quantum Annealing (QA) device, i.e. a D-Wave quantum annealer or the Fujitsu Digital Annealer Unit (DAU). A Simulated Annealing (SA) algorithm has been adopted to compute the classical baseline. We will compare the results of the implementations and propose a heuristic statement, on how to deploy QKD systems in provider networks.

We are not aware that the approach of QUBO formulation and solving it for a QKD network optimization problem has been tried out in the past. Our contributions in this paper are twofold. 
\begin{itemize}
\item We propose a novel hybrid algorithm that leverages quantum
computing methods for the optimization of QKD network
engineering solutions.
\item We provide evidence that the use of QA can improve
the overall performance when compared to traditional
optimization methods. The evaluation is based on real, but anonymized data from an Internet Service Provider (ISP) network.
\end{itemize}

\section{Related Work}
The commercial availability of quantum computers and annealers has recently created interest of industry experts searching for a more powerful way to solve hard combinatorial problems in their domain. In this paper we focus on the advancement of traffic engineering in modern communication networks, especially in quantum networks for the distribution of encryption keys. We are not aware of any work that has explored digital or quantum annealing for this specific problem yet.

Quantum computers or digital annealers indicated that they can improve the performance of finding solutions to optimization problems in the network domain. In \cite{engel} the task of optimizing the configuration of IP Routers to minimize the maximum link utilization of the network edges has been demonstrated. The publication followed a so-called two segment approach, where each data traffic between source and sink is routed via at maximum one intermediate network node. To do so, a QUBO model has been developed and optimized using digital and quantum annealing. A similar approach has been done for the Multi Commodity Flow Routing problem. 

QKD network optimization has been executed using classical methods. In \cite{gunkel} the same Deutsche Telekom network, which is the basis for this work, has been optimized using classical methods, i.e. the Minimum Spanning Tree (MST) or Single-Source Shortest Path Tree (SSSPT) algorithm, in order to ensure the key exchange between each node in the network. The authors were not confident that the results produced are in fact optimal for QKD networks and additionally faced challenges regarding limiting key rate capacity edges in their solutions \cite{gunkel}. In this work we will build upon their insights and incorporate ways to prevent those challenges. In addition, we incorporate the network traffic between the nodes in the QKD networks and also elaborate solutions with redundancies, to guarantee reliability. 

The comparison of different approaches to solve the QKD network optimization problem is extremely relevant for the architecture work on quantum networks, for example in the EuroQCI or QuNet initiatives. 

\section{Simulated Annealing}\label{sec:simualted_annealing_background}

Simulated Annealing is a popular meta heuristic in the field of optimization. It is inspired from annealing processes used in metallurgy to alter internal properties of metals, like their atom order. Related structural impurities are typically caused by cold deformation in standard metal processing. During the annealing process, the metal is heated to a material specific temperature 
and then slowly cooled down, allowing for the atoms to reorder \cite{Huang2006}. In the following, we use the common practice of reducing the meaning of ``annealing process'' to the cooling part. As this process is subject to the principle of minimum energy from thermodynamics, the atoms tend to form a crystalline structure, i.e., repairing all imperfections by minimizing the internal energy \cite{Callen2006}. 

Algorithmically, this process can be described as a search algorithm:
\begin{enumerate}
\item Start with some initial state (corresponding to the initial atom ordering at the highest temperature) 
\item Move to a neighbor of the previous state (corresponding to the atoms reordering due to their rapid movement at high temperatures)
\item Repeat 2. until the final temperature is reached.
\end{enumerate}
As the principle of minimum energy has to be accounted for in 2., worse solutions (states with a higher energy than the previous state) can effectively only be attained at high temperatures. In the standard version of simulated annealing, this is implemented by accepting worse solutions according to a certain probability, depending on the current temperature \cite{Kirkpatrick1983}:
\begin{equation}
P=\exp\left(-\dfrac{e'-e}{T}\right)
\end{equation}
Here, $P$ denotes the acceptance probability, $e$ the energy of the current state, $e'$ the energy of the worse neighbor and $T$ the temperature. This formula stems from analog dependencies in the evolution of related physical systems.

In this work Simulated Annealing is used as the classical baseline and its adaptation to the QKD optimization problem is described in Section \ref{sec:SimulatedAnnealing}.

\section{Quantum Annealing and QUBO}
Quantum Annealing is a meta-heuristic familiar to the classical Simulated Annealing and mostly known for solving optimization and decision problems \cite{Kadowaki1998}. Since quantum computing hardware increasingly maturing in the last years, there is now the possibility to not only simulate the heuristic on classical computers but exploit the quantum effects of recently developed quantum systems, for finding better and more promising solutions. D-Wave Systems are specialized in building quantum annealing hardware, which are designed to solve optimization problems. However, in order to use these systems, one typically formulates the corresponding problem in QUBO form \cite{Boros2007}, which is isomorphic to the needed input for the currently available quantum and digital annealers, i.e the Ising model.

In the following the generic QUBO formulation is shown: 
\begin{equation}\label{eq:QUBO}
\min x^tQx \qquad \text{with } x \in \{0,1\}^n 
\end{equation}
with $x$ being a vector of binary variables of size $n$, and $Q$ being an $n \times n$ real-valued matrix describing the relationship between the variables. Given matrix $Q:n \times n$, one uses the annealing process to find binary variable assignments $ x \in \{0,1\}^n$, which minimize the objective function in Eq. \eqref{eq:QUBO}.

In this work we outsource computationally intensive tasks of our proposed hybrid quantum annealing algorithm in form of a QUBO problem, in order to execute it on D-Wave Systems quantum annealing hardware. Note, that also gate based quantum computing algorithms, such as the Quantum Approximate Optimization Algorithm (QAOA) \cite{farhi2014quantum}, could be considered for the hybrid approach. However, due to the limited resources (number of qubits and their chip connectivity) of actual gate model computers, annealing based hardware was favoured.

\section{Network and Use Case}\label{sec:Network} 
\begin{figure}[ht]
 	\centering
 	\includegraphics[width=0.6\textwidth]{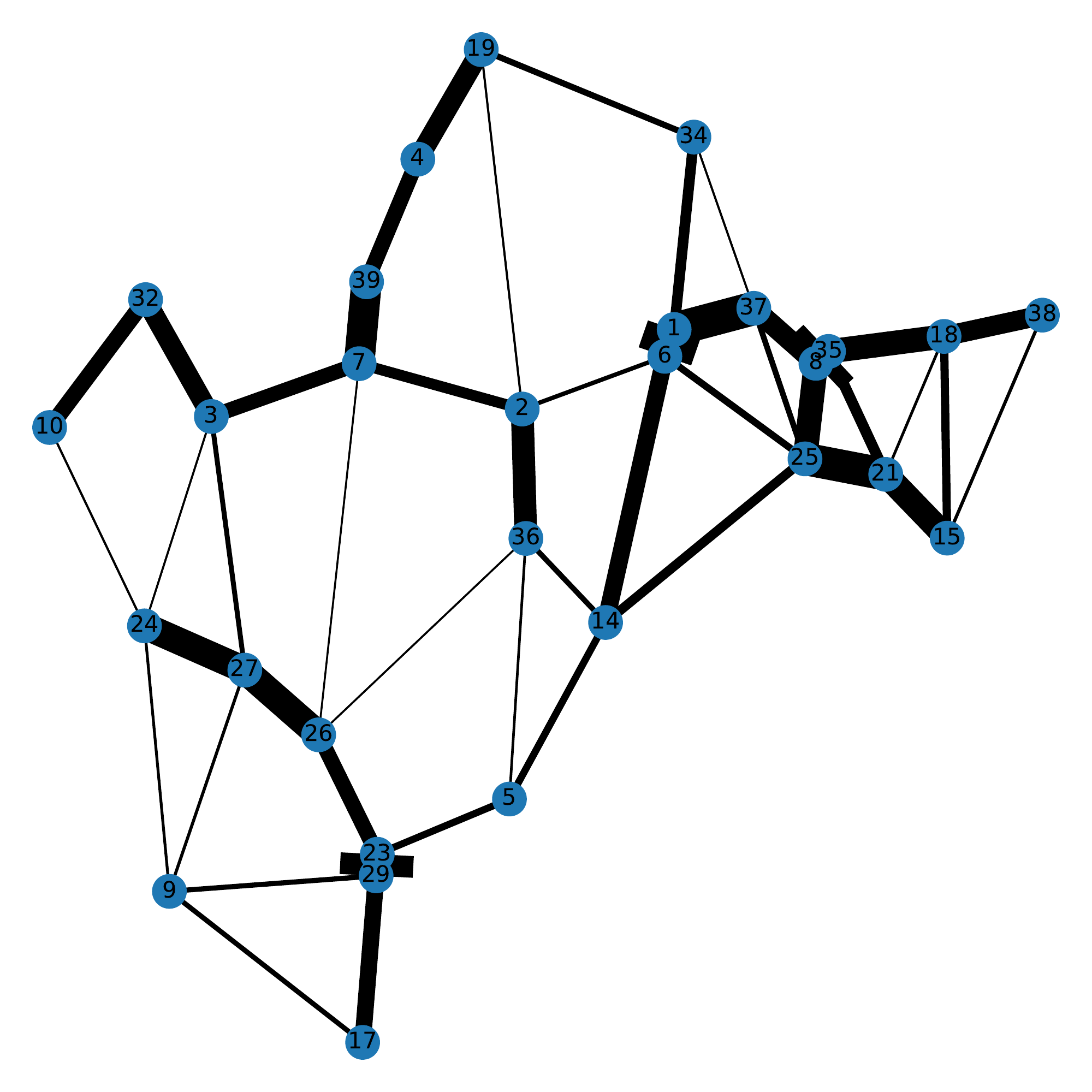}
	\caption{Network G with 29 nodes and 48 edges. The thickness of the edges correspond to their respective key rate capacity.}
 	\label{fig:network}
\end{figure}

For the evaluation of the later proposed methods an undirected reference quantum network $G=(N,E)$ consisting of 29 nodes $N$ and 48 dark fiber edges $E$ was created, see Figure \ref{fig:network}. The network, with its central backbone node $6$, corresponds exactly to the topology in \cite{gunkel}. It models possible quantum channels in a real world DT fiber network. Furthermore, w.r.t to the experiments and evaluation the geographical coordinates, the adjacency matrix, defining the dark fiber links between adjacent nodes, the capacity key rate matrix, describing the maximal possible key exchange rate of commercially available QKD systems and the traffic demand matrix, defining the amount of data to be encrypted, of the network are given.

The key rate is regarded as the measure that indicates the quantum-secure encryption capability for the data traffic involved. Thus, besides to the number of QKD systems, it can also be used as a evaluation criterion for such a network. The key rate is expressed in kbit/s, although this unit is usually omitted in the following sections for the sake of clarity.  

Within this work, we primarily focus on a so-called N:N scenario, where every network node is a source as well as a sink of the network traffic. That means, that the whole quantum-secure data transmission between all nodes must be guaranteed, i.e. the wanted QKD network must be able to exchange a sufficiently large number of encryption keys so that the traffic demand defined by the traffic demand matrix elements can be AES encrypted and transmitted via a classical transport network. The classical network does not need to be modelled in this work, however, we point out, that the routes taken for the quantum key exchange do not necessarily have to coincide with the routes taken by classical network data.

In \cite{gunkel}, a sub-problem of the N:N, the so called 1:N scenario is investigated, with a single backbone node being the source and all other nodes being sinks. The authors used exclusively 1:N algorithms, such as the MST and the SSSPT, in order to find acceptable solutions. Thereby, the problem of bottlenecks arises, i.e. there are edges in the 1:N solution, which would strongly impair the encryption capability of the data traffic of all subsequent nodes, see Figure 2 and 3 of \cite{gunkel}. Here one can see, that the purple edge restricts the subgraph circled in purple to a maximum key rate of 3.2 and 2.3 kbit/s, respectively. This suggests, that 1:N algorithms are of limited use, when also considering the traffic demand of the $N:N$ case, in which every network node is a source as well as a sink of network traffic. 

Therefore, an essential question in the N:N scenario is how to avoid the bottlenecks or how to minimize their effects. This in turn means that in addition to individual edges, connections between two non-adjacent nodes must now be taken into account within the N:N scenario. Without these connections, respectively paths, it is impossible to handle the bottleneck problem. In addition, N:N solutions with redundancies in the QKD network are investigated, in order to ensure reliability, if one link fails.   

\section{Concept}\label{sec:concept}  

\subsection{Hybrid Quantum Annealing Method}\label{sec:nn}  
Within this section we describe our Hybrid Quantum Annealing (HQA) approach for optimizing a QKD network taking the data traffic for a $N:N$ scenario into account, so that bottleneck problems in regard of the traffic are avoided. 
The quantum part is inspired by the known MST problem, for which there already exists a QUBO formulation \cite{isfor}.
However, since the existing MST QUBO formulation only considers single favourable edges in its objective function, we adapted it w.r.t also taking paths between the nodes of the network into account, in order to avoid single edges in the solution leading to bottlenecks. A detailed description of this QUBO is stated later on in subsection of \ref{sec:nnqubo}.  

Since quantum hardware is still quite limited in the number of qubits and their connectivity, we had to come up with an hybrid approach, which can tackle this $N:N$ network scenario. That means, we had to find a way to efficiently split up the overall problem into sub-problems, solve them independently, and reconstruct them to an overall solution later on. So to say, we perform N - times the adapted MST which is constrained by a path length, in order to restrict the QUBO size and be able to execute it on the limited QC hardware. 

On the classical side we came up with a discount and filter concept to reconstruct the overall solution of the single adapted MST solutions. The discount method rewards edges $uv$, that are often found in the individual MST solutions. That means, the individual adapted MST QUBOs influence each other across the iterations, so that edges which are often found in previous solutions are also more likely to appear in later solutions. Beyond that only the composition of the individual MST solutions remains. This is implemented within the filter method. For this, a descending sorting of all edges $uv$, according to their frequency in all individual solutions is proceeded. Afterwards those edges are again sorted in descending order w.r.t their key rate. Both sorts now determine the position of the respective edge in the total order. 

This allows the actual bit by bit construction of the $N:N$ solution to takes place. There is an additional minimizing condition in regard to the number of edges. This condition implies the successful termination of the construction, as soon as the $N:N$ solution is contiguous. Ideally, the solution has $|N|-1$ edges. However, it is possible that there are more edges in that solution. This is mainly due to the final sorting, since this defines the order of the edges and thus also of the construction of the overall QKD network solution. 

In Algorithm \ref{alg:n-n-overview} an overview of the hybrid approach is presented in pseudocode.
\begin{algorithm}
\caption{Hybrid Quantum Annealing Method}\label{alg:n-n-overview} 
\begin{algorithmic}[1] 
\STATE GET network $G=(N,E)$
\STATE CALCULATE pairwise paths of certain length of every node in $N$ 
\STATE CHOOSE start node $SN$ 
	\STATE{SORT $N$ after $SN$ in descending order and RETURN as $sortedN$}
	\STATE{SET up discount}
			\FOR{$sN \in sortedN$} 
				\STATE{GET paths of node $sN$}
				\STATE{CONSTRUCT QUBO}  
				\STATE{SOLVE QUBO}
				\STATE{GET solution}
				\STATE{UPDATE discount}  
		 	\ENDFOR{} 
	\STATE{FILTER and BUILD $N:N$ solution with discount}
\STATE{SAVE $N:N$ solution}
\end{algorithmic}
\end{algorithm}
Given the network $G=(N,E)$, one needs to calculate all pairs of paths between the nodes in advance, e.g. with the known depth-first-search (DFS). Due to the limited QC hardware we had to constrain the length of the paths by 6 for our reference network. In the case that some nodes were not reachable within this range we calculated the shortest path as a replacement. After that one needs to choose a root node (start node) from which the overall solution (a so called $N:N$ Minimum Spanning Tree) is constructed from. In our DT reference network this node corresponds to the core backbone node $6$. Experiments showed that in the general case it is good to choose a central network node (line 3). After choosing the start node $SN$ one sorts the other network nodes $N$ according to $SN$ in descending order from nearest to farthest (line 4). The discount function is initialized (each edge equally good) and the iterative process starts. For every node the adapted MST QUBO is constructed with not only considering the edges but also the paths of a certain length between nodes in the graph within the objective function (line 7-10). After each iteration the discount method updates and weights the edges regarding their frequency in previous solutions (line 11). We assume, that edges which are favoured in the individual solution are also advantageous in the overall solution.  
After computing all $N$ individual solutions the filter method sorts the edges according to their key rate capacity and their frequency in the computed single solutions (line 13). Afterwards the total $N:N$ solution is constructed w.r.t the order of the edges and saved (line 14).

\subsubsection{QUBO}\label{sec:nnqubo}	    
The QUBO formulation for the adapted MST can be found below. $H_A$ represents the constraints and $H_B$ the costs w.r.t the $N:N$ scenario. The constraints are composed of three partial formulations. 
\begin{equation}
\label{eq:nn}
\begin{aligned}
	H &= A\cdot H_A + B\cdot H_B \\ 
	&= A\cdot(T_1 + T_2 + T_3) +  B\cdot (\sum\limits_{vt\in P, n\neq t} c_{nt} \cdot p_{nt})	
\end{aligned}   
\end{equation} 
The condition $T1$ enforces that all nodes $n$ occur in the solution. The partial formulation $T2$ ensures that there is exactly one path or connection $p_{nt}$ from every node $n$ to every other node $t$. Thus, $T2$ ensures the minimal connectivity. Here $P$ stands for the set of all paths or links.
Finally, the condition $T_3$ implies that for each selected path $p_{nt}$, all subpaths $sec(p_{nt})$ are also included. Here, the $len(sec(p_{nt})$ represents the number of sub-paths.
\begin{equation}
\label{eq:nnha}
\begin{aligned}
  	 H_A  &= T_1 + T_2 + T_3 \\
  	 &= \sum\limits_{n\in N} (1-x_n)^2 +\sum\limits_{nt\in P, n\neq t} (x_n - p_{nt})^2 \\[-1ex]&
  	 + \sum\limits_{nt\in P, n\neq t} p_{nt} \cdot (len(sec(p_{nt}))- \sum\limits_{p_{sec} \in sec(p_{nt})} p_{sec}) 
\end{aligned}    
\end{equation} 

Moreover, the costs $H_B$ consist exclusively of the path costs $c_{nt}$, see Equation \eqref{eq:nn}, which are essentially based on the individual edge costs ($c_{nu},....,c_{qt}$). The edge costs have been significantly tightened to adequately address the previous described bottleneck problem. Therefore, the individual edge costs are defined as $c_{nu} = (1 - discount(Q_{nu}))\cdot\textit{max degree}\cdot|N|\cdot(\frac{1}{key-rate_{nu}})^2$

with \textit{max degree} being the maximum degree of any vertex in the graph and \textit{discount} describing the times an edge $nu$ has been selected in previous solutions. This number is denoted by $Q_{nu}$. Thus, the actual \textit{discount} is represented by the function $discount(Q_{nu}) = 1-(0.9^{Q_{nu}})$. In order to calculate the path costs, it is necessary to distinguish, whether there is a bottleneck or not. I.e. if the path $p_{nt}$ has a bottleneck or no bottleneck, the path costs are defined as $c_{nt} = len(p_{nt})\cdot max(c_{nu},....,c_{qt}) $ or $c_{nt} = len(p_{nt})\cdot c_{nu}$, respectively. 
The latter costs are to be interpreted that in a path $p_{nt}=n\rightarrow u\rightarrow ...\rightarrow q \rightarrow t$ with $n$ as initial node, no key-rate higher than $key-rate_{nu}$ can occur. Note, that all sub-paths of $p_{nt}$ are considered separately and the return path $p_{tn}$ is considered in the corresponding iteration, where $t$ is the root node. \\

This basic cost structure allows for the first time to actively penalize occurring bottlenecks. To further strengthen this penalty, a cost aggravating classification of the paths is introduced. After all path costs have been calculated, they are classified by a median cost classification. In total there are four classes: \textit{Worst}, \textit{Worse}, \textit{Good}, \textit{Excellent}. Those paths which belong to the classes \textit{Worst} and \textit{Worse} are penalized more, because of this classification. Furthermore, paths that belong to the classes \textit{Good} and \textit{Excellent} are favored by cost reduction. These final path costs are now given in Equation \eqref{eq:epfk}, where $c^{'}_{nt}$ is the path cost of the basic cost structure. 
\begin{equation}
\label{eq:epfk}
\begin{aligned}
	Worst &: c_{nt} = \frac{c^{'}_{nt}}{0.25 \cdot len(p_{nt})} \\
	Worse &: c_{nt} = \frac{c^{'}_{nt}}{0.5 \cdot len(p_{nt})} \\
	Good &: c_{nt} = \frac{c^{'}_{nt}}{2 \cdot len(p_{nt})} \\
	Excellent &: c_{nt} = \frac{c^{'}_{nt}}{4 \cdot len(p_{nt})}
\end{aligned}    
\end{equation} 
In total, the presented QUBO requires $|N| + |P|$ qubits, of which $|N|$ is allocated to $x_n$ and $|P| $ is allocated to $p_{nt}$.
Finally, the penalty factors $B$ and $A$ are set to $0.5$ and $(B \cdot round(2\cdot (\textit{max cost})))+1$, respectively. Here $\textit{max cost}$ stands for the maximum path cost of the present cost structure of the reference network G and $round()$ for the round-up function.

\subsection{Hybrid Quantum Annealing Method with Redundancy}
In this section the expansion regarding necessary redundancy within the HQA approach is addressed.
Redundancy is incredibly important to ensure that if an edge $nn'$ fails, the node $n$ is still reachable from every other node in the network $G$. That is, there is at least one more edge $nn''$ to allow key exchange in case of network failure. The redundancy is generally divided into separate in- and outgoing edges or paths. The goal of the expansion of the hybrid approach is to construct largest possible circles within the network, so that every node has at least one in- and outgoing edge or path and we additionally use as few as necessary edges and therefore QKD systems in the solution.   

Since this circle redundancy follows a long-range approach the consideration of paths, as in the previous approach of section \ref{sec:nn}, is indispensable. Especially because the redundancy $rd$ can be split into a separate in- and outgoing path $ip$ and $op$, respectively. These two paths underlay to the same path length constraint, as described in the previous section. This in turn means that the redundancy $rd$ has in our case a maximum length of $2\cdot 6=12$. Since it can not be guaranteed, that every node of the network is reachable within this range and therefore is part of a redundancy path, the approach is also developed in a hybrid iterative manner. This means, that it is iterated over all nodes that do not belong to at least one redundancy path at the time of the corresponding iteration.  

Furthermore, the maximum length of $rd$ implies that there are usually more redundancies than paths of length $6$ that can be found for a node $n$. I.e., the number of possible redundancies has to be restricted because of the size problem (number of qubits). Nevertheless, in order to find the largest possible redundancies, only the longest redundancies between the respective initial and target nodes are considered. These redundancies are denoted with the quantity $RD$. Furthermore, this restriction is also conducive to the goal of finding the greatest possible redundancies, since thus hardly any small redundancies can be selected. 

In addition, as with the previous approach from section \ref{sec:nn}, the question of the processing sequence arises. Since it is more effective, but not necessarily more efficient, to find the largest and best possible redundancy from a smaller set $RD$, an ascending sorting according to $RD$ has proven itself. Furthermore, an input network $IN$ in the form of a full $N:N$ solution is required to provide the minimum connectivity. 

In Algorithm \ref{alg:nnr-overview} an overview of the described hybrid quantum annealing approach with (circle) redundancy is shown. 
\subsubsection{QUBO} \label{sec:kredqubo}
The QUBO formulation for the $N:N$ scenario with redundancy can be found below. 

\begin{equation}
\label{eq:kredh}
\begin{aligned}
H &= A\cdot H_A + B\cdot H_B \\ 
&= A\cdot(T_1 + T_2 + T_3 + T_4 + T_5) +  B\cdot (C_1 + C_2 + C_3)	 
\end{aligned}    
\end{equation} 
$H_A$ represents the constraints and $H_B$ the costs of the total QUBO Hamiltonian $H$. The constraints are composed of five partial formulations. 

\begin{equation}
\label{eq:kredha}
\begin{aligned}
H_A &= T_1 + T_2 + T_3 + T_4 + T_5 \\
&=\sum\limits_{rd\in RD_{n}} (rd\cdot (len(rd)-\sum\limits_{t\in rd}r_t) + \sum\limits_{nn{'}\in E} (y_{nn^{'}} - (x_{nn'} + x_{n'n}))^2  \\[-1ex]& +(1-\sum\limits_{rd\in RD_{n}} rd)^2 + \sum\limits_{rd\in RD_{n}} rd \cdot (len(rd) \\[-1ex]& -\sum\limits_{nn',...,ab,...,un\in E} y_{nn'}+...+y_{ab}+...+y_{un}) 
\\[-1ex]& +\sum\limits_{nn'\in E_{IN} }(1-x_{nn'})^2
\end{aligned}    
\end{equation} 
In Equation \eqref{eq:kredha}, the constraint $T_1$ implies that for each selected redundancy $rd$ the corresponding redundancy nodes $r_t$ are set, with $len(rd)$ representing the length of the redundancy. The constraint $T_2$ states that each undirected edge $y_{nn'}$ occurs exactly once and thereby each directed edge $x_{nn'}$ respectively $x_{n'n}$ occurs exactly once. The constraint $T_3$ ensures that exactly one redundancy $rd$  is selected. The constraint $T_4$ adds the single undirected edges $y$ for that redundancy. Finally, constraint $T_5$ implies the minimum connectivity by adding all edges from the input network $IN$. 
		 
\begin{equation}
\label{eq:kredhb}
\begin{aligned}
H_B &= C_1 + C_2 + C_3 \\
&=\sum\limits_{nn'\in E} c_{nn'} \cdot x_{nn'} +  \sum\limits_{rd\in RD_t} c_{rd} \cdot rd + \sum\limits_{n\in N} c_{r} \cdot r_n
 \end{aligned}    
\end{equation} 

In addition, the total cost Hamiltonian $H_B$ consists of the edge costs $C_1$, the redundancy costs $C_2$, and the redundancy node costs $C_3$. The edge costs $C_1$ are defined analogously to the previous approach in the section \ref{sec:nn}. The redundancy cost $C_2$ and in particular $c_{rd}$ are derived from the round-trip costs. I.e., $c_{rd} = c_{ip}+c_{op}$, where the path costs $c_{ip}$ and $c_{op}$ are formulated analogously to \ref{sec:nnqubo}, including the median cost classification of the paths. 
The last part of $H_B$ consists of the costs of the redundancy nodes $C_3$. These are determined by $c_{r} = \sqrt{\frac{min(c_{rd})}{(5\cdot len(N))^6}} $, with $min(c_{rd})$ representing the minimum redundancy cost. 

The justification for $C_3$ is found in the fact, that the longer the redundancy, the more often potentially cost-increasing edges are added. This means, that the input network $IN$ or all previous redundancies do not contain these edges. Overall, $C_3$ has only a slight cost-increasing effect, since the focus is mainly, but not exclusively, on the longest possible redundancy.		
Overall, the (circle) redundancy QUBO presented here requires exactly $|OP| + |N| + 3 \cdot |E|$ qubits, of which $|OP|$ is allocated to $op$, $|N| $ to $r_{n}$, $|E|$ to $y_{nn'}$, $2\cdot |E|$ to $x_{nn'}$.

Finally, the penalty values $B$ and $A$ are set to $0.5$ and $(B \cdot round(2*\textit{max cost}))$, respectively. Here $\textit{max cost}$ stands for the maximum redundancy cost of the present cost structure of the reference network and $round()$ for the round-up function. 

\subsubsection{Bridges} \label{sec:bridges}		
The so-called bridge problem comprises the circumstances, that two strictly separated networks may occur if an edge has been removed from the original network.
Therefore it is a compelling condition that in particular the (circle) redundancy supplies solutions, which do not contain any bridges. Otherwise the redundancy criterion that in case of failure or removal of any edge, all nodes are still connected to each other, would be obsolete.

An interesting conjecture directly tangent to the bridge problem was independently postulated as the so-called \textit{Cycle Double Cover Conjecture} in \cite{sey,sze}. Adapted to redundancy, this aims at the fact, that it seems to be sufficient to solve the bridge problem, if each node is at least part of two different redundancies. Since this conjecture is still considered open to date and the implementation would lead to too large QUBO formulations, a different approach is proposed below. 

However, it should be noted that, depending on the input network, it is possible not to achieve complete redundancy. In this context, incomplete redundancy means, that not every node belongs to at least one redundancy. However, for the present reference network it is possible to achieve that completeness consistently with this concept. 

\begin{figure}[htbp]
 \centering
 \includegraphics[width=0.6\textwidth]{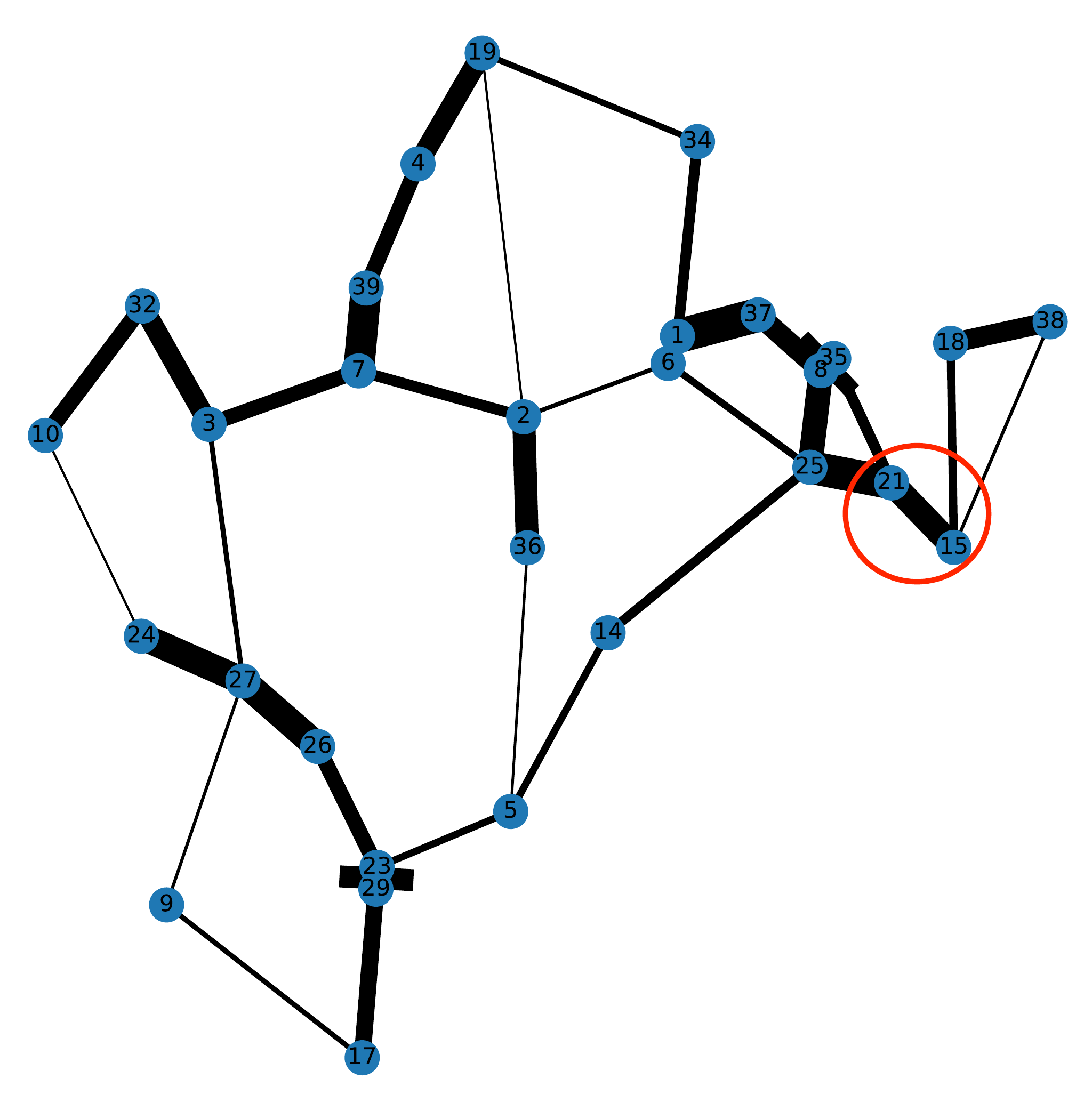}
 \caption{Example $N:N$ solution incl. redundancies and a bridge}
 \label{fig:bri2}
\end{figure}

The basic idea behind the concept is stated in form of an example and becomes clear if one looks at a $N:N$ solution of the reference network including redundancy and a bridge, as stated in Figure \ref{fig:bri2}.  

\begin{example}
One can clearly see the bridge, corresponding to the edge $(21,15)$. If this edge is removed a larger and smaller network occurs. The latter is composed of nodes $15$,$18$ and $38$. None of these three nodes can have a connection or redundancy with the larger one. At the same time, the concept starts with this property. We propose as a workaround to recompute the redundancy completely, where the nodes $15$,$18$ and $38$ can only belong to a redundancy that contains at least one node from the larger network. Referring to our example, $35$ would be such a node. 
\end{example}

This workaround is executed and run until either a defined upper bound for the purpose of computation time limitation is reached or the circle redundancy no longer contains any bridges. Finally, this concept does not require any further qubits, but only computing time.

\begin{algorithm}[htbp]
\caption{Hybrid Quantum Annealing Method with Redundancy}\label{alg:nnr-overview}
\begin{algorithmic}[1]
\STATE GET input network $IN=(N,E)$
\FOR{$n \in N$}
	\STATE{GET and SAVE paths}
	\STATE{SEARCH, RETURN and SAVE redundancies} 
\ENDFOR{} 
\STATE{GET all redundancies}
\STATE{SORT $N$ after $|redundancies|$ in ascending order and RETURN as $sortedN$}
\STATE INITIALIZE $bridge$ with False 
\WHILE{\NOT $bridge$} 
	\STATE{INITIALIZE redundancy nodes set $rns = \emptyset$} 
	\FOR{$n \in sortedN$}
		\IF{$n \notin rns$} 
			\STATE{CONSTRUCT QUBO} 
			\STATE{SOLVE QUBO}
			\STATE{RETURN redundancy solution as $rs$}
			\STATE{UPDATE $rns$ with redundancy nodes from $rs$} 
			\STATE{UPDATE input network $IN$ with $rs$}
		\ENDIF 
	\ENDFOR{}  
	\STATE{RETURN $IN$ as complete redundancy solution $crs$}
	\IF{$crs$ has bridge}
		\STATE{DETECT all bridges}
		\STATE{WORKAROUND all bridges}
	\ELSE
		\STATE{RETURN $crs$ as bridgeless complete redundancy solution $bcrs$}
		\STATE{SET $bridge$ to True}
	\ENDIF 
\ENDWHILE{} 
\STATE{SAVE $bcrs$}
\end{algorithmic}
\end{algorithm}

\section{Evaluation}
\subsection{Solver Methods}
\subsubsection{D-Wave Hybrid Solver}\label{sec:SolverDW} 
For solving the QUBO problems in the two proposed iterative hybrid quantum methods the D-Wave System Hybrid Solver was used, which was introduced in the 2020 launch of Leap 2 as part of the Hybrid Solver Service \cite{leap2}. The basic configuration of the Hybrid Solver consists exclusively of the maximum computing time, since all other parameters, which mainly affect the QPU, are set internally. The maximum computing time is set to $10s$. Indeed, field tests showed that the default maximum computing time of $3s$, was not sufficient in this QKD optimization scenario.

Furthermore, the Hybrid Solver allows a maximum problem size of $10.000$ qubits with a resolution of 4-bit \cite{dwprec}. Again, field tests have shown, that the limit for reasonable or valid solutions (in this use case) is just under $1.000$ qubits. I.e. for our problem formulations, QUBO instances with sizes between $1.000$ and $10.000$ variables over strain the QPU, respectively Hybrid Solver.  

\subsubsection{Classical Simulated Annealing}\label{sec:SimulatedAnnealing} 
In order to use the previously described generic Simulated Annealing algorithm (see, Section \ref{sec:simualted_annealing_background}) for the QKD optimization use case, several quantities have to be assigned:
\begin{enumerate}
\item The energy (objective function): With respect to the N:N use case scenario, the optimization target is twofold. On the one side, we desire the highest possible connectivity $r$ for all nodes, and on the other side, the number of needed QKD systems $qkd$ should be minimal.
That is why, we set $r$ to the sum of the ratio $\frac{t}{c}$ of the traffic $t$ to the capacity $c$ of each edge lying on a shortest path. Field testing revealed $e=r+100 \cdot qkd$ as a useful assignment.
\item Neighbor generation: For this, a straight forward probabilistic approach was used. First select an arbitrary subset $E'$ of edges from the input graph containing at least two edges at random. Then, to construct a neighbor, delete all edges from the current candidate that are in $E'$ and insert all edges that are missing from the current candidate, but are contained in $E'$. This process gets repeated until the resulting graph is connected. In case redundancy is demanded, its verified that the graph is still connected after each possible single edge deletion.
\item The annealing schedule: 
\begin{itemize}
\item Start temperature: 1000
\item Final temperature: 0.0001
\item Step size, i.e. the evolution of temperature $t\rightarrow \alpha t$, where $\alpha=0.9$ and $t$ is the temperate of the respective iteration.
\end{itemize}
\end{enumerate}
In addition, a slight adaptation of the standard acceptance probability by inserting a term accounting for the number of times $w$ no new candidate solution has been accepted, was made:
\begin{equation}
P=\exp\left(\dfrac{e-e' + \beta w}{T}\right)
\end{equation}
The scaling factor $\beta$ was set to 0.04 based on field testing. This adaption essentially allows to escape local minima easier, when otherwise getting stuck.

\subsection{Results}\label{sec:Results} 
\subsubsection{1:N Scenario}
Even so the scope of this work was mainly on the investigation of the N:N scenario, we also reviewed the 1:N case, as in \cite{gunkel}. The authors stated that finding a Minimum Spanning Tree in the reference network is a valid approach to address the needs of QKD network design for such a scenario. However they were not sure, if their MST results computed with classical methods were optimal. Therefore we used a simple QUBO model for the MST as stated in \cite{isfor} to verify their results by using D-Wave's quantum annealing hardware.

However, without going into detail, the results weren't any better and exhibit the same bottlenecks edges as in \cite{gunkel}. That is why we summarize and assume that for the respective reference network those limiting edges can't be avoided.

\subsubsection{N:N Scenario}
In the following, the N:N results obtained by using D-Wave Hybrid Solver within the Hybrid Quantum Annealing method compared to the classical Simulated Annealing approach are described. The edge improvement and minimum key rate were used as evaluation criteria. The edge improvement describes the number of saved edges compared to the reference input network and is calculated by $\frac{edges_{total}-edges_{solution}}{edges_{total}}$, with $edges_{total}$ being the number of edges of the reference network and $edges_{solution}$ the number of edges in the solution. Thus the edge improvement is an indirect measure for the number of QKD systems saved. The minimal key rate states the edge in the solution with the lowest key rate capacity and is therefore a measure for possible capacity bottlenecks. 

\begin{figure}[H]
\centering
\subfloat[]{\label{fig:a_QA}\includegraphics[width=0.47\linewidth]{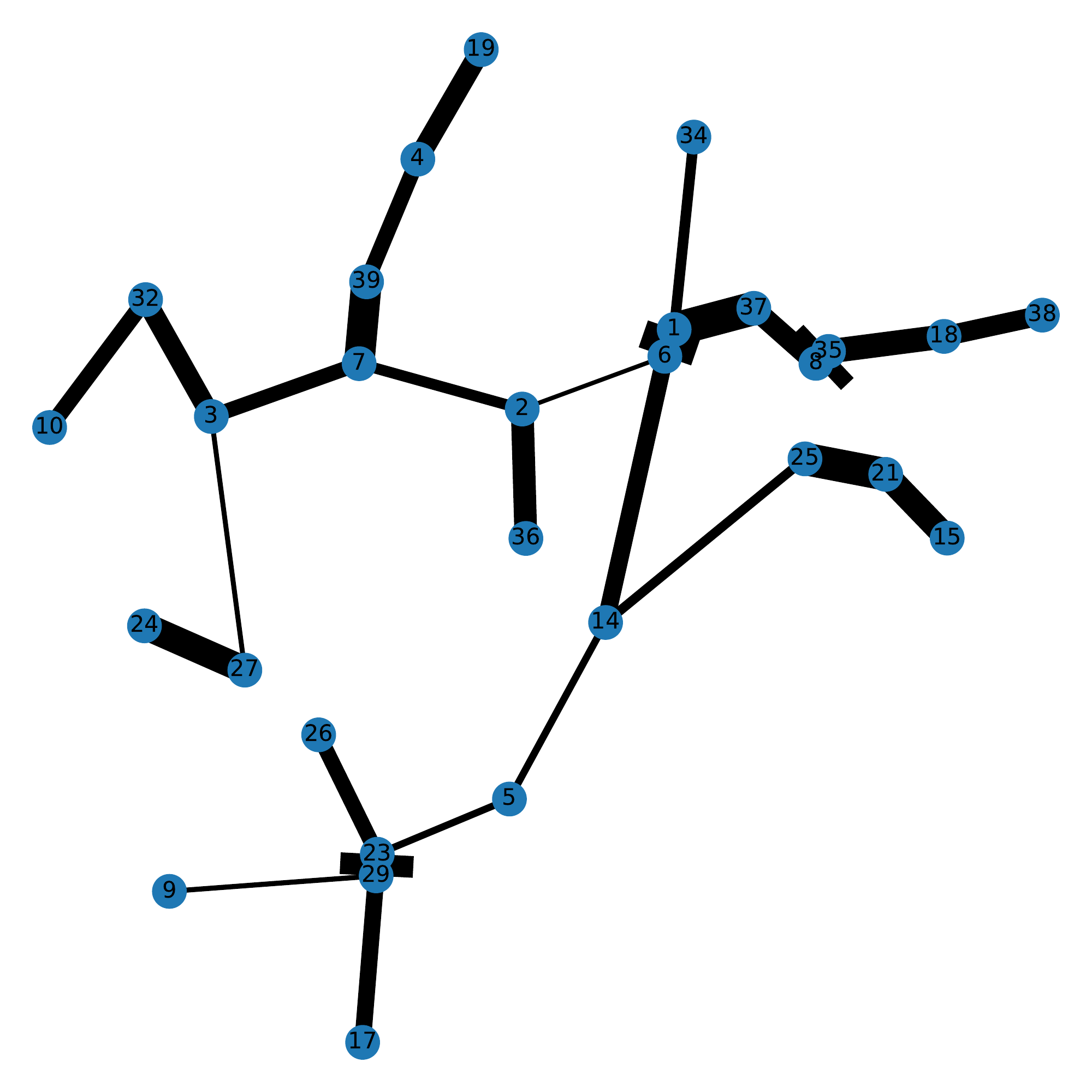}}\qquad
\subfloat[]{\label{fig:b_QA}\includegraphics[width=0.47\linewidth]{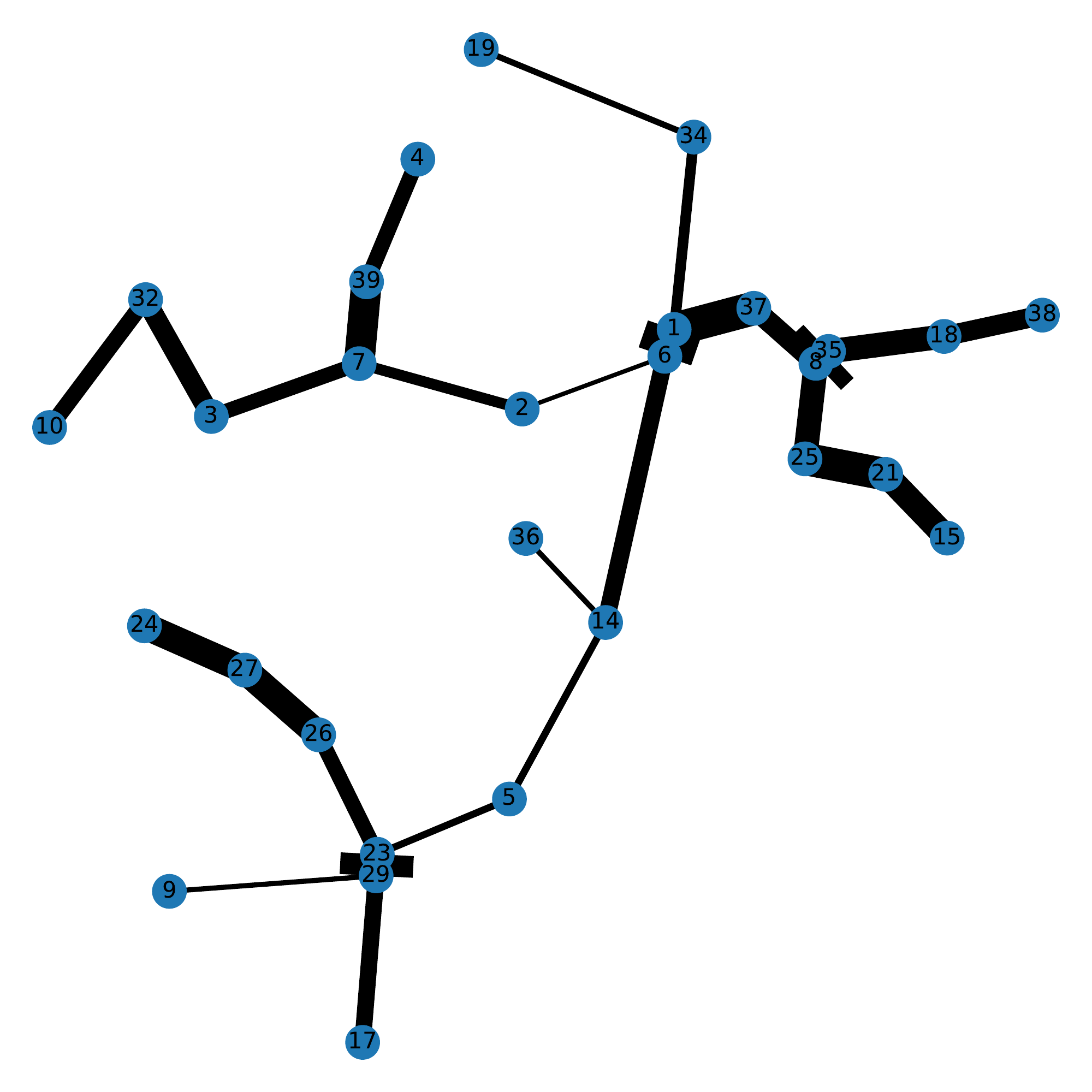}}\\ 
\subfloat[]{\label{fig:c_QA}\includegraphics[width=0.47\textwidth]{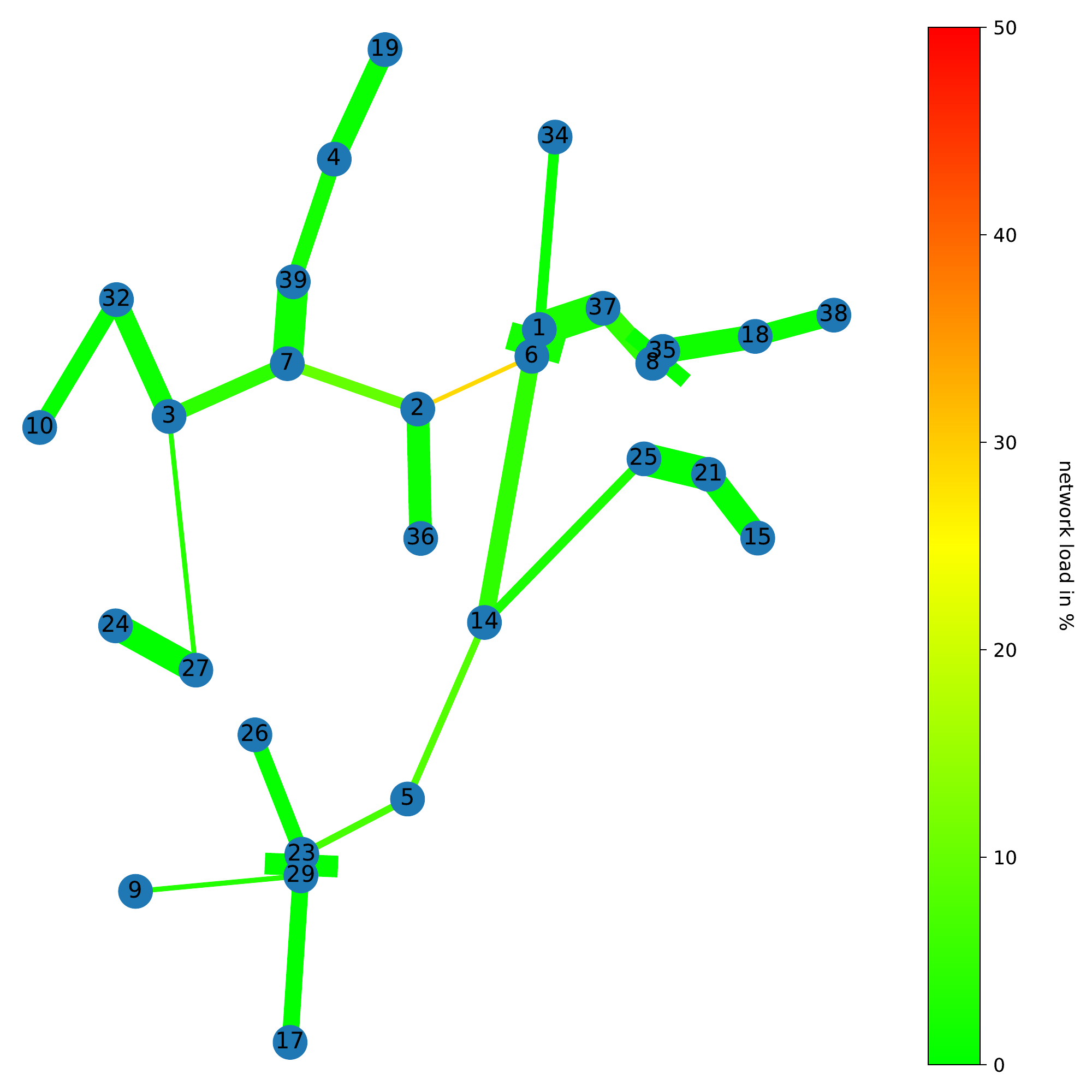}}\qquad
\subfloat[]{\label{fig:d_QA}\includegraphics[width=0.47\textwidth]{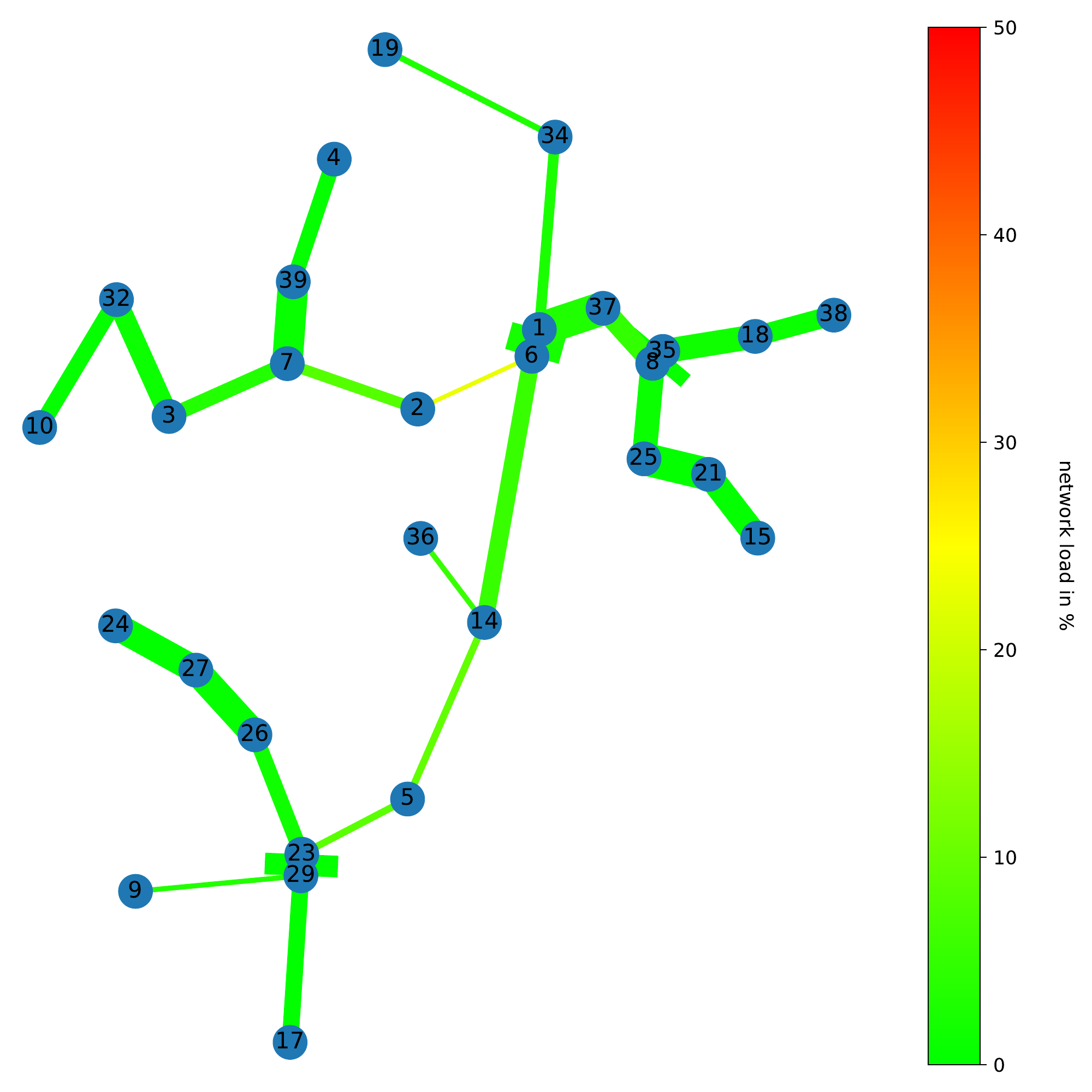}}
\caption{
In (a) the best Hybrid Quantum Annealing N:N solution with an edge improvement of 41.67\% and a minimal key rate of 2.27 kbit/s is shown, while in (c) its traffic workload is plotted. 
In (b) the best found Simulated Annealing N:N solution with an edge improvement of 41.67\% and a minimal key rate of 2.27 kbit/s is shown, while in (d) its traffic workload is plotted.}
\label{fig:NN_result_plots}
\end{figure}

In Figure \ref{fig:NN_result_plots}, the best found solutions for HQA and SA for the N:N scenario without redundancy are shown. One can see that both solution methods can reach the global optimum of an edge improvement of 41.67\% with a minimal key rate of 2.27 kbit/s. Even so the plotted best solutions for both methods (see Figure \ref{fig:NN_result_plots} (a) and (b), respectively) are different, the solution quality w.r.t the minimal key rate and edge improvement is the same.

For determining the network load in the corresponding solutions traffic plots were generated. They describe the workload of each edge of the solution when taking the traffic matrix into account. 
For the N:N scenario without redundancy the traffic plots of the HQA and classical SA solution (see Figure \ref{fig:NN_result_plots} (c) and (d), respectively), contain both the limiting edge between node 2 and 6 with a minimal key rate of 2.27 kbit/s. However, the sub-graph of the SA solution is more favourable, since fewer nodes have to be supplied via this limiting edge. In \cite{gunkel} SSSPT found a similar solution for the 1:N case, with the same limiting edge between node 2 and 6. That is why we assume that either the solutions in general requires that limiting edge of the reference network in order to create a superior overall N:N solution, or all three methods (HQA, SA, SSSPT) get stuck in that local optimum. 
  
\begin{figure}[H]
\centering
\subfloat[HQA - N:N without redundancy]{\label{fig:a_NN-EI}\includegraphics[width=0.3\linewidth]{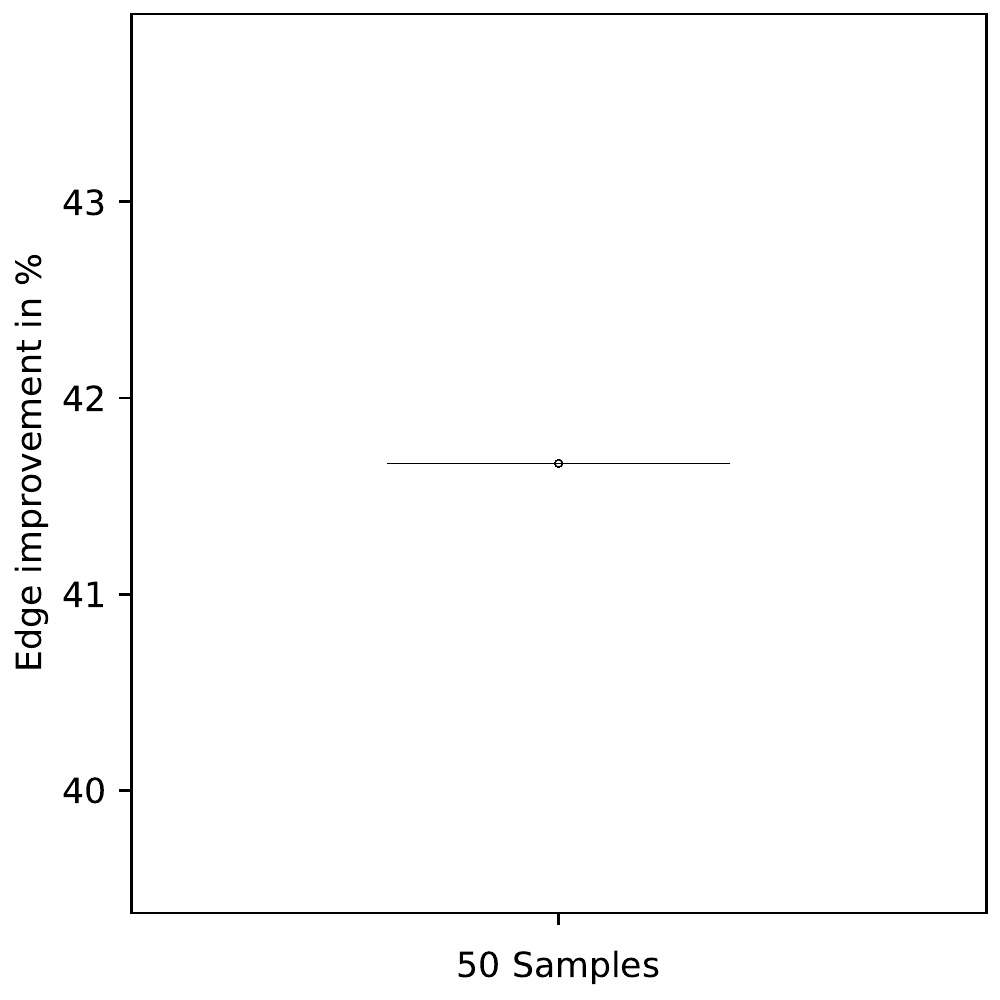}}\qquad
\subfloat[SA - N:N without redundancy]{\label{fig:b_SA-NN-EI}\includegraphics[width=0.3\linewidth]{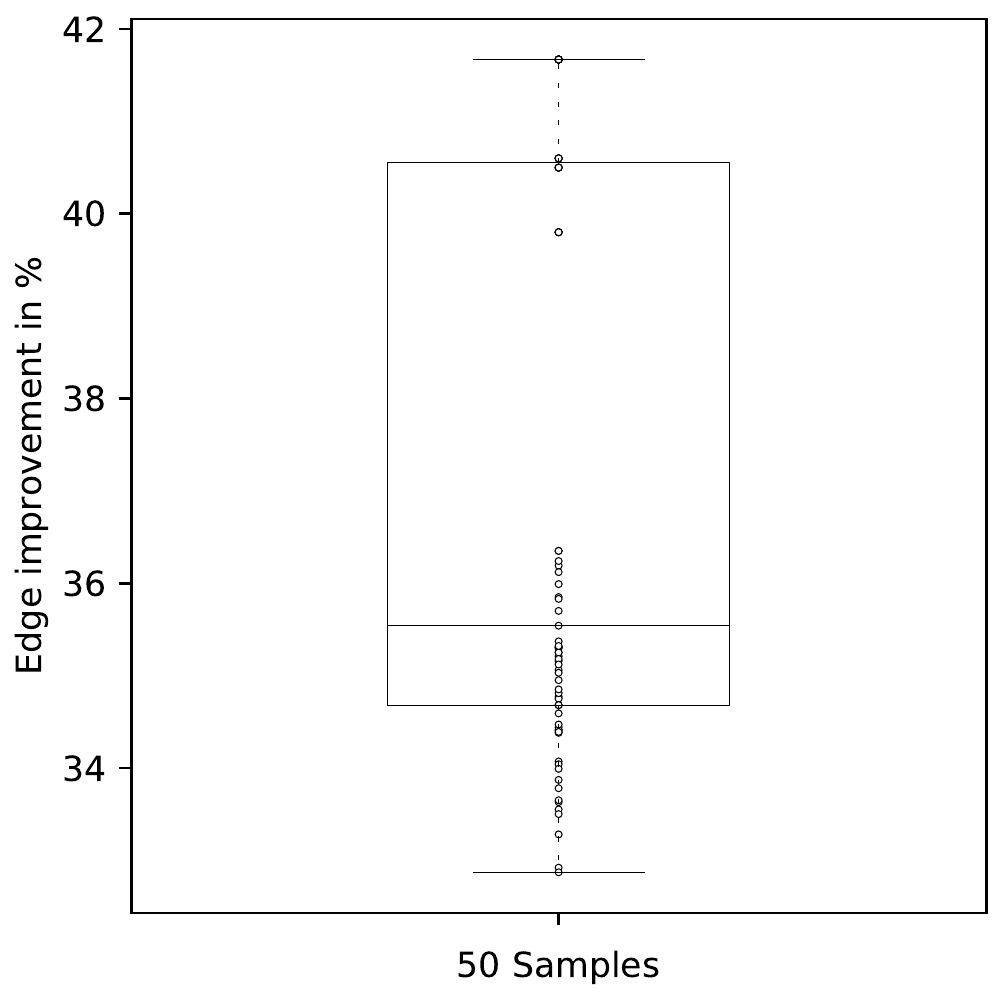}}\\
\caption{
Best edge improvement of HQA and SA N:N solutions out of 50 runs.
}
\label{fig:QA_SA_NN_boxplots}
\end{figure}

In Figure \ref{fig:QA_SA_NN_boxplots} the edge improvement of each best solution out of 50 runs is represented in a box-plot. Since our HQA approach consists of many adapted MST solutions, which are in the end constructed to an overall N:N solution, the best possible edge improvement of 41.67\% is mostly guaranteed within a run. The SA results show a larger variance regarding the edge improvement, with an average of 34.8\% and an optimum of 41.67\%. 

\begin{figure}[htbp]
\centering
\subfloat[]{\label{fig:a_QA_red}\includegraphics[width=0.47\linewidth]{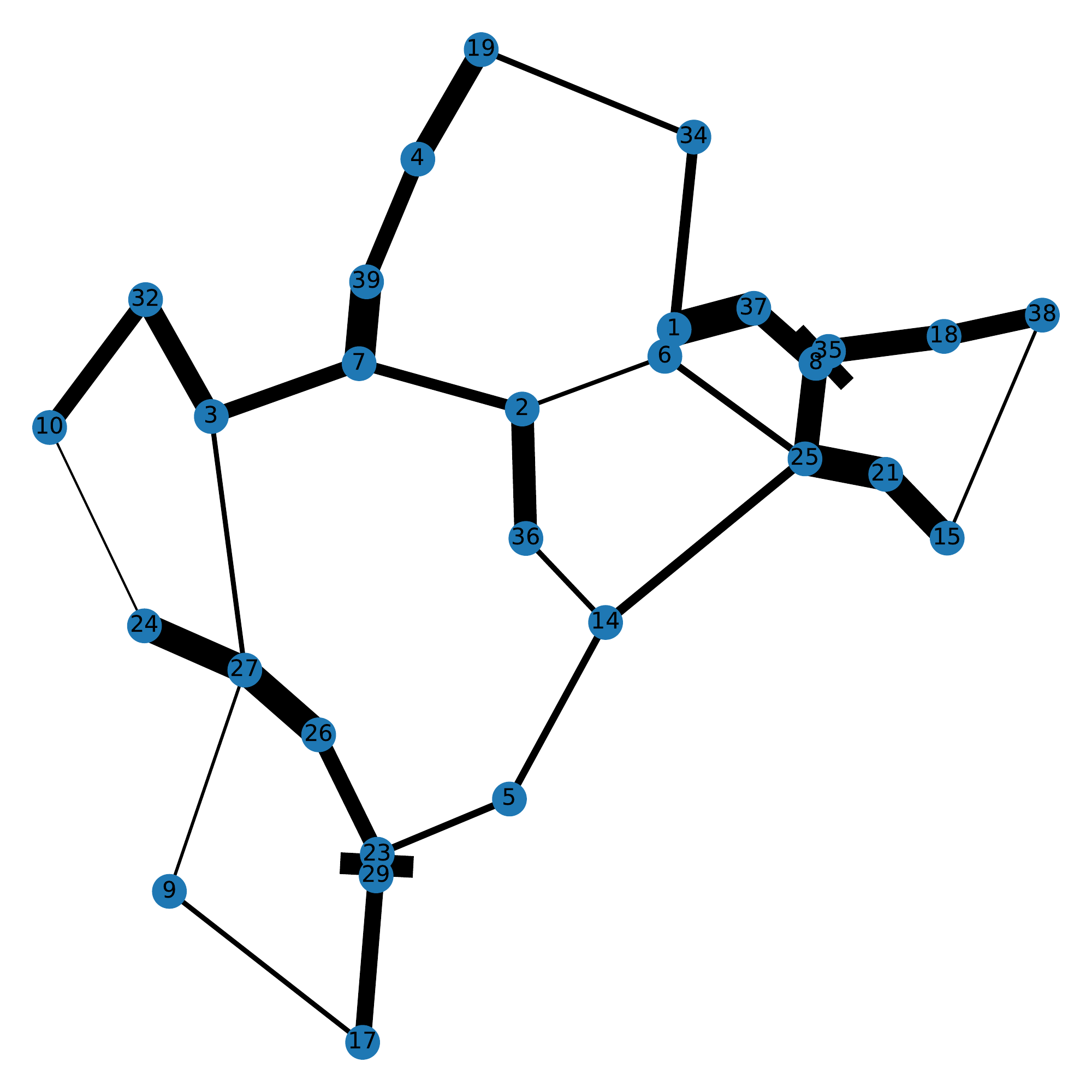}}\qquad
\subfloat[]{\label{fig:b_SA_red}\includegraphics[width=0.47\linewidth]{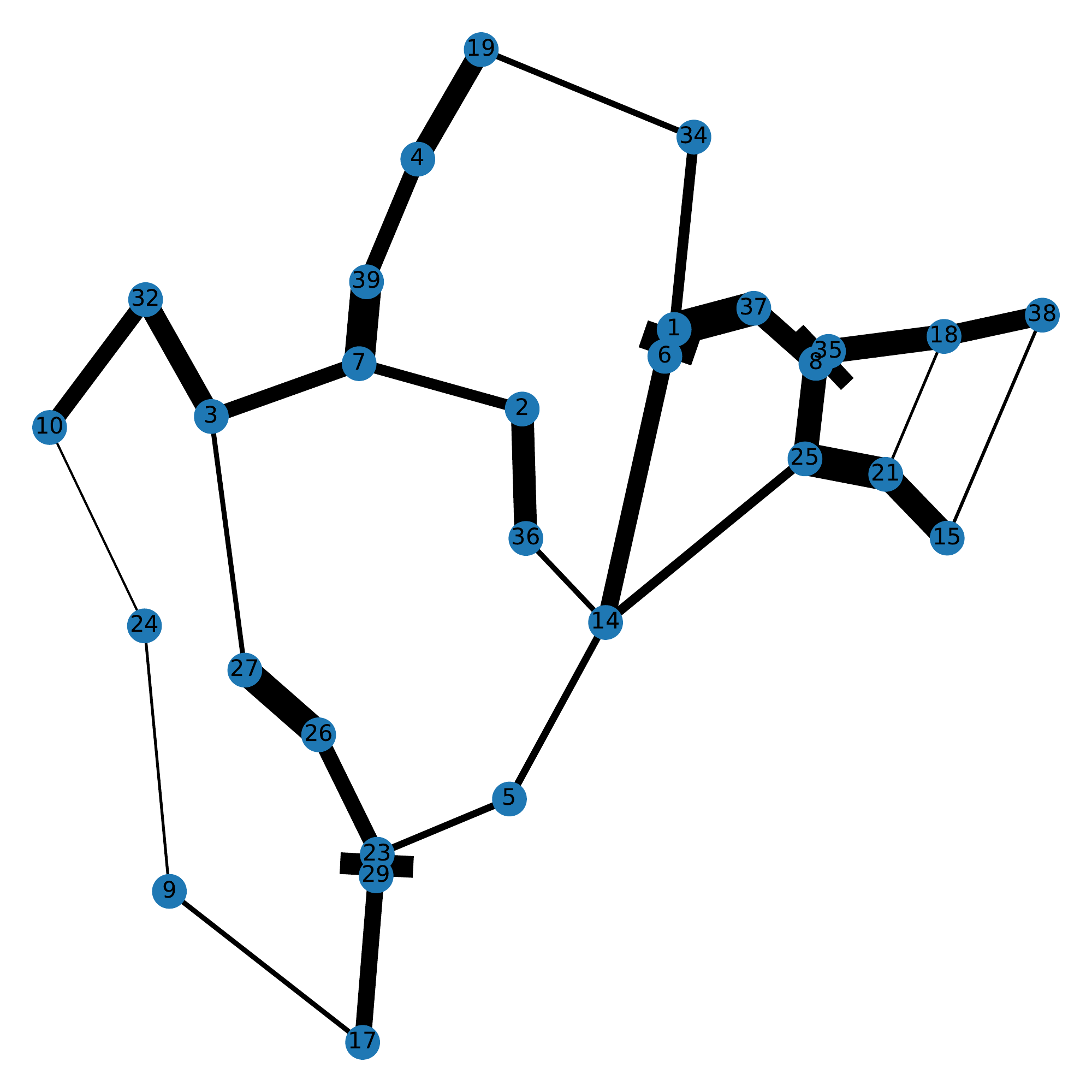}}\\
\subfloat[]{\label{fig:c_QA_red}\includegraphics[width=0.47\textwidth]{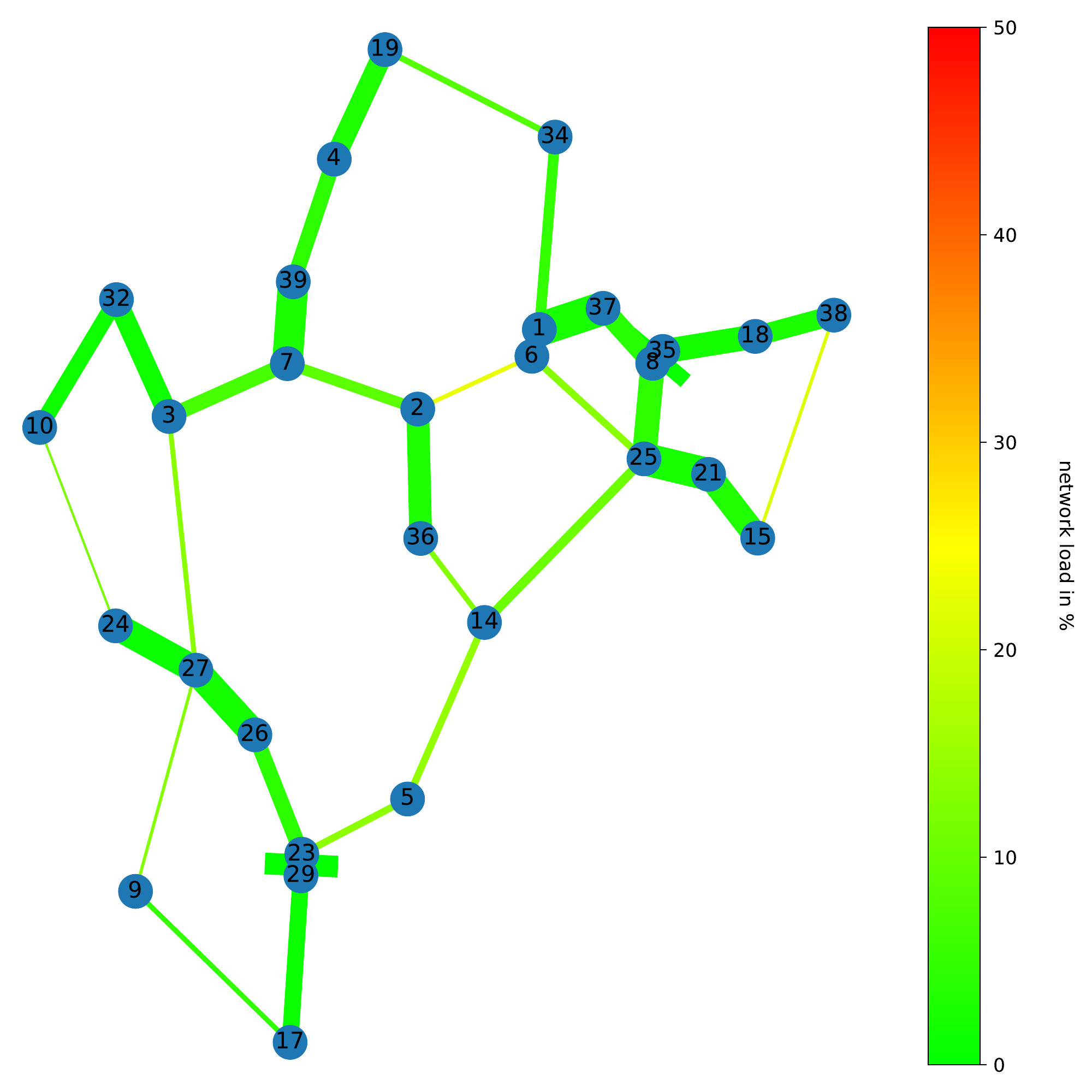}}\qquad
\subfloat[]{\label{fig:d_SA_red}\includegraphics[width=0.47\textwidth]{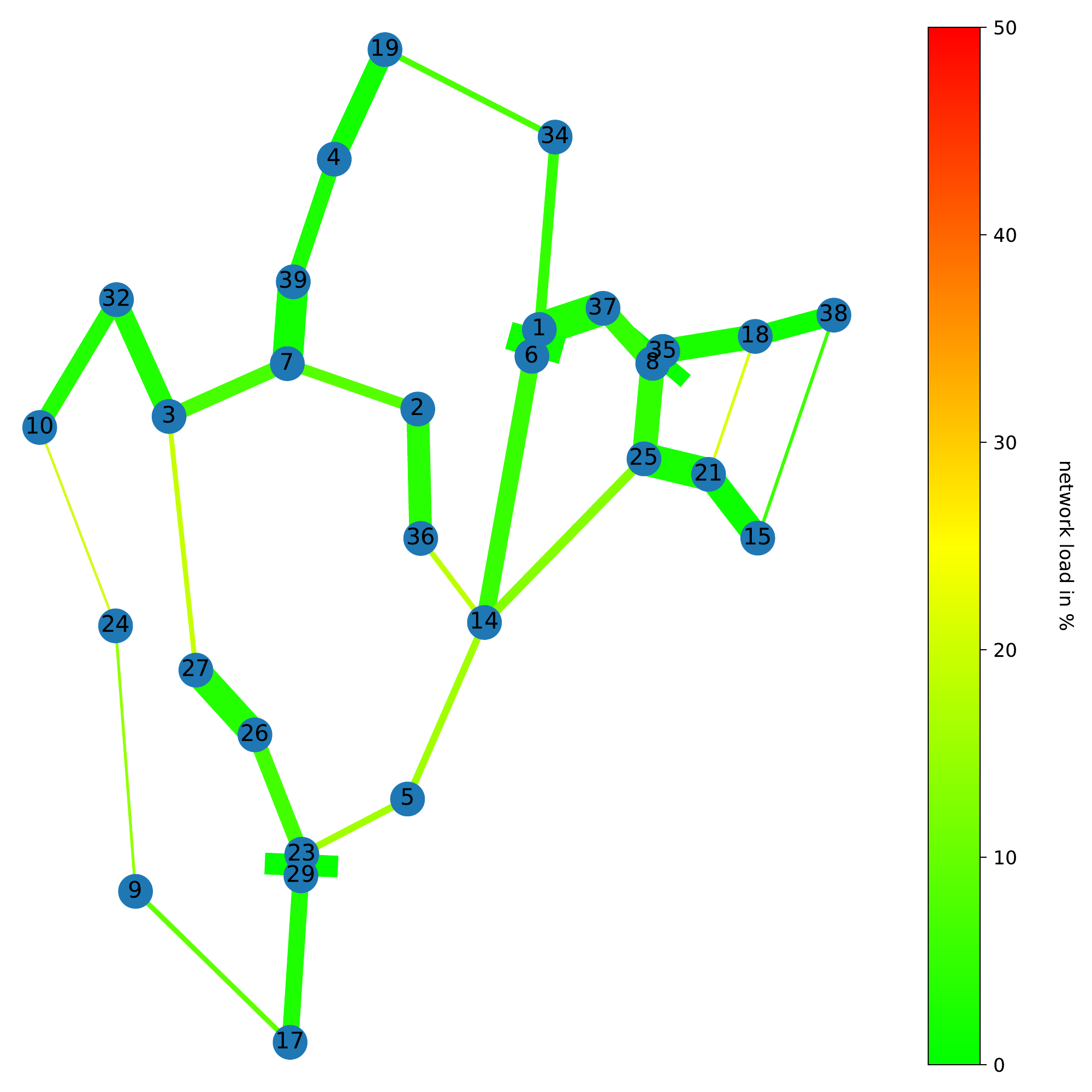}}
\caption{
In (a) the best Hybrid Quantum Annealing N:N solution including the circle redundancy with an edge improvement of 29.17\% and a minimal key rate of 1.14 kbit/s is shown, while in (c) its traffic workload is plotted. 
In (b) the best Simulated Annealing N:N solution including the redundancy with an edge improvement of 29.17\% and a minimal key rate of 1.14 kbit/s is shown, while in (d) its traffic workload is plotted.}
\label{fig:NN_redundancy_result_plots}
\end{figure}

In Figure \ref{fig:NN_redundancy_result_plots}, the best found solutions for HQA and SA for the N:N scenario with (circle) redundancy are shown. Again, each solution methods are equal regarding the assessment criteria and find solutions with an edge improvement of up to 29.17\% and a minimal key rate of 1.14 kbit/s. 
However, the best solutions found by HQA and also SA are not the global optimum. While in the HQA solution the edge between node 2 and 6 could be neglected, in the SA solution, the edge between node 18 and 21 is not necessary. Computational results showed that by manually neglecting those edges the overall fitness of both solutions would be better. However, both methods keep getting stuck in these local optima. 

Within the N:N scenario with (circle) redundancy the network load was intensified, see traffic plots in Figure \ref{fig:c_QA_red} and \ref{fig:d_SA_red}. In this context, we simulate how high the load on an edge $k$ would be, if any edge failed. All edges, except for $k$, are allowed to fail one after the other, in order to determine the highest load for $k$. The results show, that due to the redundancy, for both solutions the network load was at a maximum of approximately 27.0\%. So to speak the key exchange capacity of nowadays QKD systems is by far sufficient to encrypt the data payload of the DT reference network. The average network load does not exceed 30\% on any edge, which gives enough room to build up an encryption key backlog or to compensate operation failures.

\begin{figure}[htbp]
\centering
\subfloat[HQA - N:N with circle redundancy]{\label{fig:c_NN-CR-EI}\includegraphics[width=0.3\textwidth]{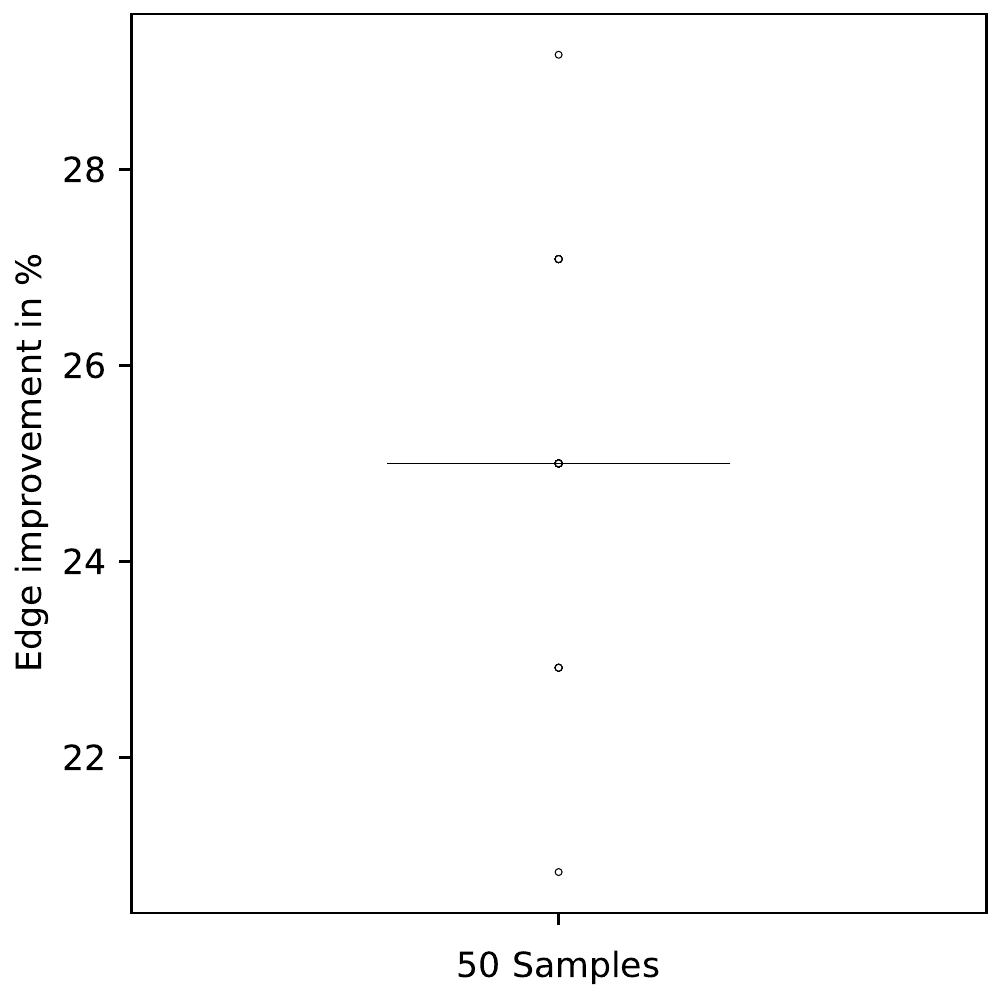}}\qquad%
\subfloat[SA - N:N with redundancy]{\label{fig:d_SA-NN-R-EI}\includegraphics[width=0.3\textwidth]{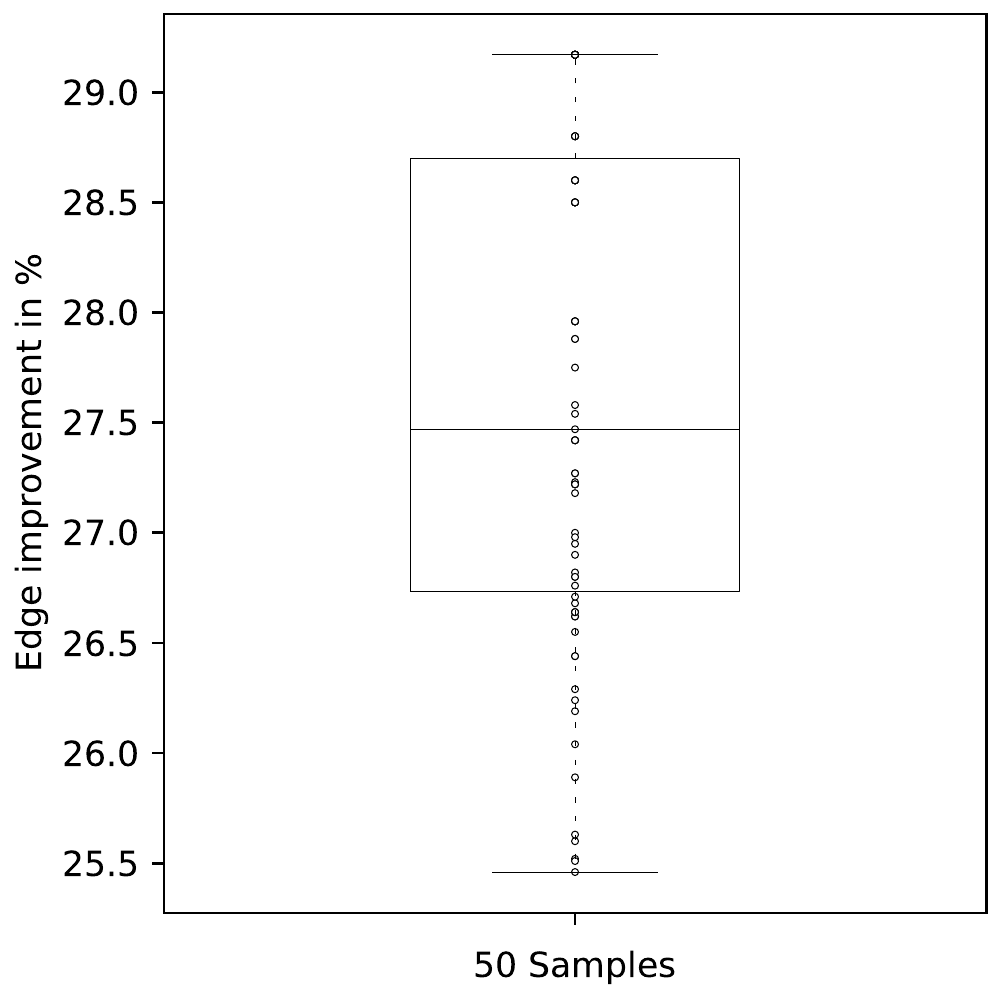}}\\
\caption{
Best edge improvement of HQA and SA N:N solutions with (circle) redundancy out of 50 runs.}
\label{fig:QA_SA_NN_wR_boxplots}
\end{figure}

In Figure \ref{fig:QA_SA_NN_wR_boxplots} the edge improvement of each best solution out of 50 runs is represented in a box-plot. The average edge improvement of the HQA approach is 25.0\%, while the average edge improvement of the SA solutions is around 27.4\%. However both methods found an equal local optimum with a 29.17\% edge improvement.    

Both methods found equally good solutions for the N:N scenario with and without redundancy. While in the former 20 edges could be saved in the latter 14 edges could be neglected and still ensure reliability, if one link fails. Those savings have an enormous influence on the deployment of QKD systems in the reference network. In general for each saved edge a pair of QKD systems of the cost of around 200.000\EUR{} can be avoided.  

However, regarding the runtime of both methods, the SA algorithm is superior. The HQA approach contains large classical pre- and postprocessing phases (path computation, reconstruction of individual MST solutions to an overall solution, the QUBO creation etc.), which makes it unfavorable to use for the investigated DT reference network. Nevertheless, we believe, that with respect to more complicated and key-rate capacity restricted networks the computational intensive HQA might deliver more beneficial solutions, than the SA method, since it takes paths into account in order to avoid capacity bottlenecks. 

\subsection{Heuristic Statement}\label{sec:Heuristic} 
The computed results of the optimization problem indicate a classical heuristic that can be used to plan an efficient QKD network with redundancy. The heuristic is shown in Figure \ref{fig:obersavtion_plot} and consists of the following steps:

\begin{figure}[htbp]
\centering
\includegraphics[width=0.98\textwidth]{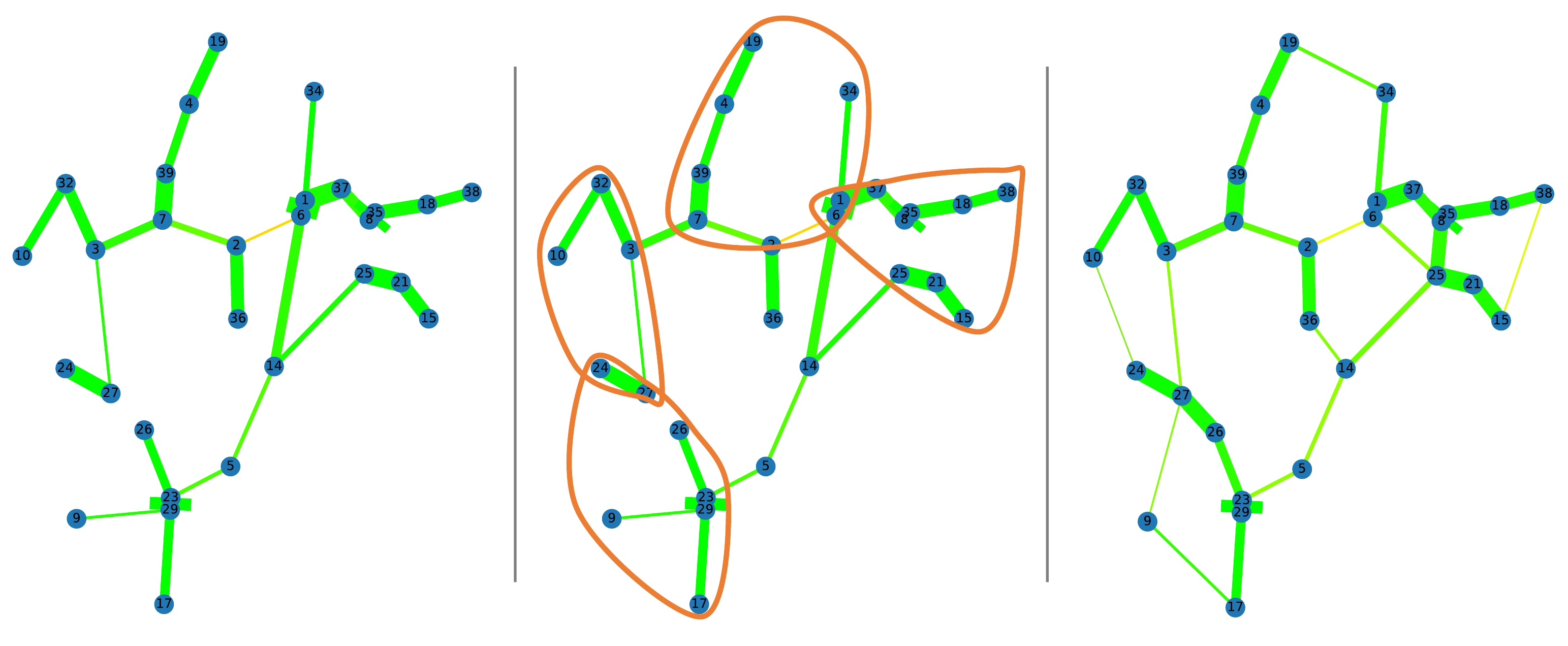}
\caption{
Left: showing the computed solution of the optimized QKD network without redundancy. Right: showing the computed solution of the optimized QKD network with redundancy. Middle: indicating a classical heuristic to derive the solution with redundancy (right) from a solution without redundancy (left) by activating network links forming large circles around the satellite nodes.
}
\label{fig:obersavtion_plot}
\end{figure}
\begin{enumerate}
    \item{Compute the MST solution of the network starting from a central network node, i.e. the Internet Peering node or a core network node using a classical computer. The reasoning is that it has been demonstrated before that MSTs are optimal results for the problem without redundancy. The MST solution is shown on the left side of Figure \ref{fig:obersavtion_plot}.}
    \item{Find adjacencies in the network that connect the satellite nodes of the MST solutions forming biggest circles possible. These circles ensure redundant solutions, because each node on the circle is connected by two sides. If more than one circle is possible, the circle with the largest key exchange capacity needs to be chosen. All satellite nodes of the MST need to be included within one circle each. This step is indicated by the red circles in Figure \ref{fig:obersavtion_plot}.}
    \item{Add the new adjacencies to the quantum network, i.e. deploy additional QKD systems to the network end points of the new network links, to obtain a redundant solution.}
\end{enumerate}

The right side plot of Figure \ref{fig:obersavtion_plot} shows the solution of the QKD network installation in the example DT network including redundancy using the method described in this section. The result is close to the computed solution of Figure \ref{fig:NN_redundancy_result_plots}. 
\section{Conclusion}
This paper successfully applied the methods of classical and quantum annealing to the problem of QKD network optimization, i.e. to find the most cost effective deployment of QKD systems in a network to be able to encrypt the forcasted payload data between all nodes. The major learnings of the computations, w.r.t to the reference network of DT, are:

\begin{itemize}
    \item{Simulated and Quantum Annealing are methods that are well suited to compute the optimal deployment of QKD systems in a network with regards to the installation sizing while maximizing the overall key exchange rate in the network. Quantum Annealing is still limited by the hardware with its small number of qubits, their connectivity and the suboptimal resolution of the control circuits.} 
    \item{The Minimum Spanning Tree is the optimal solution for a non redundant QKD network. QKD systems should be installed at every network node along the MST.}
    \item{Adding redundancy, i.e. asking for at least two key exchange paths between all the network nodes, the annealing algorithms are able to compute an optimal solution with regards to the installation sizing and overall key exchange rate. Redundancy requires additional links to be equipped with QKD systems.}
    \item{From the results of the computation, it is possible to derive a classical heuristic to find the optimal QKD network installation including redundancy. This is done by adding the biggest possible circles to the result of the MST computation, followed by adding existing network adjacencies to connect the MST satellite nodes along the circles.}
    \item{The key exchange capacity of nowadays QKD systems is by far sufficient to encrypt the data payload of the DT example networks. The average load rates do not exceed 30\% on any edge, which gives enough room to build up an encryption key backlog or to compensate operation failures.}
\end{itemize}

\subsubsection*{Acknowledgements.}
This work was funded by the German BMWK project PlanQK (01MK20005I and 01MK20005L).

\bibliographystyle{splncs04}
\bibliography{bibliography}

\end{document}